\documentclass[twocolumn]{aastex631} 
\usepackage{amsmath}

\newcommand{\target}{PSR~B0540-69}
\newcommand{\oneE}{1E~2259+586}

\definecolor{darkgoldenrod}{rgb}{0.72, 0.53, 0.04}

\definecolor{dca}{rgb}{0.64, 0.0, 0.0}
\definecolor{pp}{rgb}{0.7, 0.0, 0.9}

\begin{document}

\title[\target{} anti-glitch]{Discovery of the first anti-glitch event in the rotation-powered pulsar PSR B0540-69}

\correspondingauthor{Y. Tuo, M. Ge, and A. Santangelo}
\email{tuo@astro.uni-tuebingen.de, \\ gemy@ihep.ac.cn, \\ santangelo@astro.uni-tuebingen.de}

\author[0000-0003-3127-0110]{Youli Tuo}
\affiliation{Institut f\"{u}r Astronomie und Astrophysik, Kepler Center for Astro and Particle Physics, Eberhard Karls Universit\"{a}t T\"{u}bingen, Sand 1, 72076 T\"{u}bingen, Germany}

\author[0000-0002-0155-334X]{Muhammed Mira\c{c} Serim}
\affiliation{Institut f\"{u}r Astronomie und Astrophysik, Kepler Center for Astro and Particle Physics, Eberhard Karls Universit\"{a}t T\"{u}bingen, Sand 1, 72076 T\"{u}bingen, Germany}

\author[0000-0002-5470-4308]{Marco Antonelli}
\affiliation{CNRS/in2p3, Laboratoire de Physique Corpusculaire, 14050 Caen, France}

\author[0000-0002-9989-538X]{Lorenzo Ducci}
\affiliation{Institut f\"{u}r Astronomie und Astrophysik, Kepler Center for Astro and Particle Physics, Eberhard Karls Universit\"{a}t T\"{u}bingen, Sand 1, 72076 T\"{u}bingen, Germany}
\affiliation{ISDC Data Center for Astrophysics, Universit\'e de Gen\`eve, 16 chemin d'\'Ecogia, 1290 Versoix, Switzerland}

\author[0000-0002-5470-4308]{Armin Vahdat}
\affiliation{Institut f\"{u}r Astronomie und Astrophysik, Kepler Center for Astro and Particle Physics, Eberhard Karls Universit\"{a}t T\"{u}bingen, Sand 1, 72076 T\"{u}bingen, Germany}

\author[0000-0002-3776-4536]{Mingyu Ge}
\affiliation{Key Laboratory of Particle Astrophysics, Institute of High Energy Physics, Chinese Academy of
Sciences, 19B Yuquan Road, Beijing 100049, China}
\affiliation{University of Chinese Academy of Sciences, Chinese Academy of Sciences, Beijing 100049, People's Republic of China}

\author[0000-0003-4187-9560]{Andrea Santangelo}
\affiliation{Institut f\"{u}r Astronomie und Astrophysik, Kepler Center for Astro and Particle Physics, Eberhard Karls Universit\"{a}t T\"{u}bingen, Sand 1, 72076 T\"{u}bingen, Germany}

\author[0000-0002-0105-5826]{Fei Xie}
\affiliation{Guangxi Key Laboratory for Relativistic Astrophysics, School of Physical Science and Technology, Guangxi University, Nanning 530004, China}


\begin{abstract}
Using data from the Neutron star Interior Composition ExploreR (NICER) observatory, we identified a permanent spin frequency decrease of $\Delta\nu=-(1.04\pm0.07)\times 10^{-7}\,\mathrm{Hz}$ around MJD\,60132 in the rotation-powered pulsar \target{}, which exhibits a periodic signal at a frequency of $\nu\sim 19.6\,\mathrm{Hz}$. This points to an anti-glitch event, a sudden decrease of the pulsar's rotational frequency without any major alteration in the pulse profile or any significant increase of the pulsed flux. Additionally, no burst activity was observed in association with the anti-glitch. To date, observations of the few known anti-glitches have been made in magnetars or accreting pulsars. This is the first anti-glitch detected in a rotation-powered pulsar. 
Given its radiatively quiet nature, this anti-glitch is possibly of internal origin. Therefore, we tentatively frame this event within a proposed mechanism for anti-glitches where the partial `evaporation' of the superfluid component leads to an increase of the normal component's moment of inertia and a decrease of the superfluid one.
\end{abstract}


\keywords{Neutron stars(1108) -- Pulsars(1306) -- Rotation powered pulsars(1408) -- Compact objects(288)}

\section{Introduction} 
\label{sec:intro}

Isolated pulsars are neutron stars (NSs) characterized by stable rotational dynamics, with periods that span from milliseconds to several tens of seconds \citep{manchester2016vizier, enoto2019observational}. They exhibit long-term spin-down due to angular momentum loss via electromagnetic radiation, together with possible contributions from relativistic particle outflows which form the pulsar wind nebula, and the emission of gravitational waves \citep{hobbs2010international, aasi2015narrow}. 
Rotation-powered pulsars (RPPs) are a subset of isolated pulsars characterized by emissions primarily driven by magnetic braking. In contrast, another subset of isolated pulsars, magnetars, power their emission largely from the decay of their strong magnetic fields \citep[e.g.,][]{kaspi2017magnetars, Mereghetti2015}. 
However, the distinction between RPPs and magnetars is blurred because some peculiar high B-field RPPs -- objects with intermediate properties between the two classes -- are known~\citep{kaspi2011AIPC,harding2013,borghese2023review}.


The spin-down of isolated neutron stars is not completely regular and predictable, and two kinds of timing instabilities are often observed: glitches and timing noise~\citep{dAlessandro1996,Danjela2019Noise}.
Glitches, commonly observed in RPPs, are sudden increases in the rotation frequency of a pulsar often followed by an exponential relaxation~\citep[see][for recent reviews on the glitch phenomenon]{haskell2015models, Antonelli2022, antonopoulou2022pulsar, zhou2022pulsar}. 
Conversely, there are few observed events where the overall glitch contribution to the pulsar's rotation frequency is negative, referred to as `anti-glitch' or `spin-down glitch'. 

%

So far, most observed glitches are spin-ups that are instantaneous to the timing data's accuracy, interrupting the otherwise regular spin-down of a canonical pulsar or a magnetar~\citep[e.g.,][]{espinoza2011study,yu2013detection, Serim2017glitch, basu2022jodrell}. 
On the other hand, anti-glitch events have been observed only in magnetars or in accreting pulsars. \citet{archibald2013anti} discovered a sudden decrease in the rotational frequency in magnetar \oneE. Moreover, in binary pulsars where accretion from the companion star primarily governs the long-term spin evolution, anti-glitch events have been observed in NGC 300 ULX-1 \citep{ray2019anti}. 
The same internal mechanism invoked for glitches observed in spinning down pulsars -- the angular momentum transfer from the internal superfluid component to the normal one \citep{anderson1975pulsar} -- can result in an `anti-glitch' in accreting neutron stars that are spinning up~\citep{ducci2015properties,howitt2022MNRAS,Antonelli2022}. 
However, even after the seminal work of \citet{anderson1975pulsar}, it is not obvious which internal mechanism could lead to an anti-glitch in spinning down RPPs.
Apart from this theoretical difficulty, the absence of radiatively quiet anti-glitches in RPPs sparks a debate about whether RPPs can abruptly spin down without a change in their external braking torque. This work presents the first anti-glitch in an RPP, the pulsar~\target{}.

The young rotation-powered pulsar \target{} (also known as PSR~J0540-6919) was discovered in the early 1980s by the \textit{Einstein} X-ray Observatory \citep{seward1984discovery}. It is located in the Large Magellanic Cloud, surrounded by a pulsar wind nebula, and has a characteristic age of $\sim$1700 years \citep{petre2007x}. 
With a spin period of $\sim$\SI{50}{\milli\second}, \target{} shows typical characteristics of young pulsars, particularly in its timing behaviour (it is sometimes referred to as one of the `Crab's twins'). 
In 2011, \target{} experienced a sudden change in spin-down rate \citep{marshall2015discovery}, followed by the enhancement of the flux of pulsar wind nebula by 32\% \citep{ge2019brightening}. 
The braking index evolution of \target{} has been extensively studied \citep{marshall2016new, wang2020braking}, and several spin-up glitches were discovered in \textit{Rossi} X-ray Timing Explorer data~\citep{ferdman2015long}. 

Our paper is structured as follows. Section \ref{sec:observation} details the X-ray observations and data reduction. Section \ref{sec:timing} presents the timing analysis of \target{}, including the glitch model solution. The analysis of flux variations is presented in Section \ref{sec:flux}. Finally, in Section~\ref{sec:discussion}, we discuss and tentatively interpret our findings in the light of the anti-glitch models of \cite{yim2024continuous}, \cite{kantor2016} and~\cite{kantor2014anti}. 


\section{Observation and data reduction}
\label{sec:observation}

\target{} was observed using NICER. NICER's X-ray Timing Instrument (XTI) is equipped with 56 X-ray `concentrator' optics (XRC) paired with silicon drift detectors (SDD), each accompanied by a Focal Plane Module to collect X-ray photons \citep{gendreau2016neutron}. NICER/XTI offers excellent timing stability, maintaining a root-mean-square (RMS) of 100 nanoseconds relative to Universal Time \citep{enoto2021enhanced}. This precision is particularly advantageous for pulsar timing monitoring and other timing analyses. 

Our study encompasses the entire publicly available NICER/XTI dataset for \target{}, spanning from March 14, 2018, to December 25, 2023 comprising a total of 145 observations. We perform the standard Level 2 data generation processes by using the NICER pipeline tool \texttt{nicerl2} implemented in \texttt{HEADAS} (v6.31.1). Two instrumental parameters are applied to exclude the electron precipitation-type flare: an overshoot rate of <20\,counts/s, and a Cutoff Rigidity (COR\_SAX) of >1.5\,GeV. 
The barycentric correction for each photon is applied using the \texttt{barycor} tool implemented in \texttt{HEADAS}. The solar ephemeris JPL.DE430 is utilized. The precise celestial coordinates of \target{}, with a right ascension of $05^{\mathrm{h}}40^{\mathrm{m}}10.84^{\mathrm{s}}$ and a declination of $-69^{\circ}19'54.2''$ according to SIMBAD Astronomical Database \footnote{https://simbad.u-strasbg.fr/simbad/} is employed~\citep{brown2021gaia}.

Following the recommended GTI filtering criteria, the good time interval (GTI) was empty for some observations near the glitch epoch (ObsID: 6020010132, 6020010133, 6020010134, and 6020010139). Therefore, we used the Level 1 data for their timing analysis.
This approach ensured maximal data utilization, as data outside the valid GTIs, despite their substantial background and particle events, do not significantly impact the timing behavior and pulse profile shape due to their random distribution (see Section \ref{sec:timing} for details). 
Our analysis includes photons in the energy range of 0.5--12\,keV for the timing and pulsed flux measurements.


\section{Results}
\label{sec:results}

\subsection{Timing}
\label{sec:timing}

We performed the timing analysis for the whole public NICER archive. Here, we only present data around the anti-glitch epoch, from MJD\,59901 to MJD\,60303. 
The time of arrival (ToA) for each NICER observation is obtained via a phase-coherent analysis. As usual, the ToAs and their uncertainties are calculated by cross-correlating each pulse profile with a long-term cumulative profile template~\citep{taylor1992pulsar, huppenkothen2019stingray}. 
Before the glitch epoch, the long-term evolution of the ToAs in the phase domain can be approximated by a truncated Taylor series,
\begin{equation}
    \label{eq:phi}
    \Phi(t)= \Phi_0 + \nu(t-t_0) + \dfrac{\dot{\nu}}{2}(t-t_0)^2 + \dfrac{\ddot{\nu}}{6}(t-t_0)^3,
\end{equation}
where $\nu$, $\dot{\nu}$, and $\ddot{\nu}$ are the spin frequency, frequency derivative, and second derivative of frequency at a reference time $t_0$ before the glitch, respectively. 
The timing model before the glitch is sufficient to describe the spin-down trend of the pulsar, with an RMS residual of \SI{737.259}{\micro\second}. 
The values for each parameter are listed in Table~\ref{tab:spinparameters}.

\begin{figure*}[ht!]
    \centering
    \includegraphics[width=1.0\textwidth]{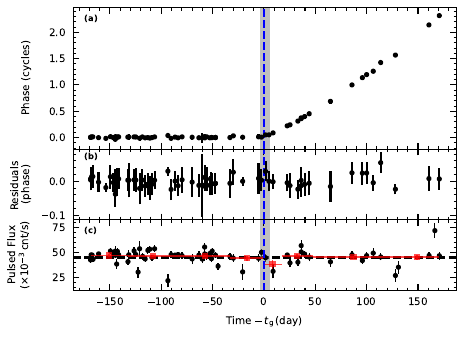}
    \caption{Timing results and the pulsed flux evolution of the radiatively quiet anti-glitch of \target{}. The inferred epoch $t_g=\mathrm{MJD}\,60132$ is marked with a blue dashed line, and the shaded band around it represents its uncertainty. 
    \textbf{(a)} The phase evolution, based on the pre-glitch polynomial ephemeris. The pulsar's phase exhibits a linear deviation from the pre-glitch solution. 
    \textbf{(b)} The residuals are fitted with the glitch model, which incorporates the terms for permanent frequency and frequency derivative increments. 
    \textbf{(c)} The pulsed flux variation around the glitch based on NICER observations in the 0.5--12\,keV range. Here, black circles represent individual observations, while red squares denote averages compiled over time.
    }
    \label{fig:phase}
\end{figure*}

As shown in the top panel of Figure \ref{fig:phase}, at around MJD\,60132 the ToAs exhibited a sudden deviation to the timing model, which indicates a glitch event. To fit the ToAs, the assumed model for the post-glitch phase residues is~(cf.~equation~\ref{final})
\begin{equation}
    \label{eq:glitch}
    \begin{split}
    &\Delta \Phi(t) =    
    \Delta \nu \, \delta t + \frac{\Delta\dot{\nu}}{2} \delta t^2 + 
    \Delta\nu_{d} \,\tau \left[ 1 - e^{-\delta t/\tau} \right]   
    \\
     & \delta t= t-t_g >0\, ,    
    \end{split}
\end{equation}
where $t_g$ is the time when the glitch/anti-glitch occurs, $\Delta\nu$ and $\Delta\dot\nu$ are the permanent changes in frequency and frequency derivative.
The above expression is valid for $\delta t\geq 0$, while for $\delta t<0$ we assume that the spin-down is well described by the phase model in \eqref{eq:phi}. An exponential post-glitch recovery term is often required in glitch modelling, where $\tau$ is the recovery time scale when a transient frequency increment $\Delta\nu_{d}$ decays exponentially to zero~\citep[see, e.g.,][]{wong2001observations,wang2012ApSS,yu2013detection}.  

\begin{figure*}
    \centering
    \includegraphics[width=1.0\textwidth]{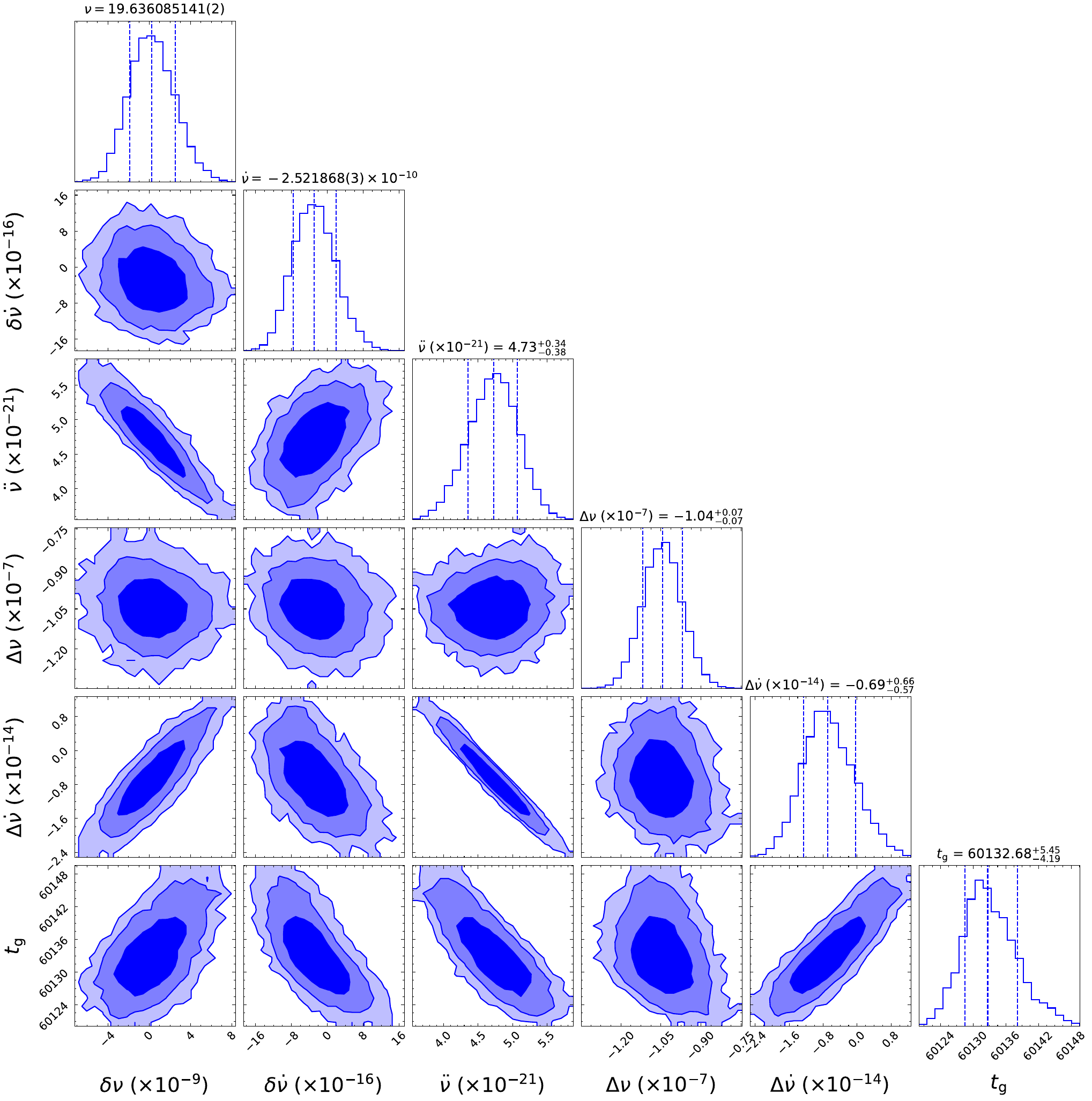}
    \caption{
    Corner plot for the posterior probability distributions of the glitch model parameters. In the two-dimensional projections, the contours for each pair of parameters denote the 1, 2, and 3-$\sigma$ credible intervals. 
    The marginal distributions for each parameter are shown along the diagonal:
    the median values of each parameter are shown, alongside the 68\% credible intervals. 
    The labels $\delta\nu$ and $\delta\dot{\nu}$ indicate that we shift the posterior distributions for $\nu$ and $\dot{\nu}$ by their median posterior values, which are exactly the ones reported in Table~\ref{tab:spinparameters} for $\nu$ and $\dot{\nu}$.
    }
    \label{fig:mcmclinear} 
\end{figure*}

We use Markov Chain Monte Carlo (MCMC) techniques, as implemented in the \texttt{emcee} package \citep{foreman2013emcee}, to fit ToAs and determine the optimal parameters for the timing model in equation \eqref{eq:glitch}.  
Details on the priors are given in Appendix~\ref{app4par}. We choose a Gaussian likelihood -- see, e.g., equation (15) of \cite{montoli2020bayesian} for the explicit likelihood's formula -- and employ 16 walkers for the MCMC. We find that 320,000 steps for each walker are enough for the statistical convergence of our Bayesian fit of \eqref{eq:phi} and \eqref{eq:glitch}. The posterior probability distributions and the marginal distributions for each parameter are shown in Figure~\ref{fig:mcmclinear}. 

Unfortunately, our Bayesian analysis indicates that the exponential terms in equation~\eqref{eq:glitch} cannot be reliably constrained. This ambiguity arises from the circumstance that the decay timescale is extremely long, leading to a reduction to a linear term that contributes to the permanent increment in frequency, or a case of rapid recovery over a short timescale, where $\tau=1.06_{-0.79}^{+1.22}\,\mathrm{day}$). 
The F-test indicates that the inclusion of the frequency derivative increment is favoured over a model that incorporates only a single term for frequency increment. 
This conclusion is supported by an F-statistic value of 17.6146 and a probability of $9.4\times 10^{-5}$. 

The inferred parameters of \eqref{eq:glitch}, along with the timing ephemeris for the spin-down model in \eqref{eq:phi}, are reported in Table~\ref{tab:spinparameters}.
The uncertainty associated with each parameter is represented by the 68\% credible interval derived from the posterior distribution.
The RMS of timing residual after the overall fit is $\SI{829.8}{\micro\second}$. The negative value of $\Delta\nu=-1.04_{-0.07}^{+0.07}\times10^{-7}\,\mathrm{Hz}$ indicates the anti-glitch nature of this event, see the discussion in Section~\ref{sec:spinup}.

\begin{table}[ht]
\centering
\begin{tabular}{lc}
\hline\hline\textbf{Parameters} & \textbf{Values} \\ \hline
    R.A. (J2000) & $05^{\mathrm{h}}40^{\mathrm{m}}10.84^{\mathrm{s}}$ \\
    Decl. (J2000) & $-69^{\circ}19'54.2''$ \\
    $\nu$ (Hz) & 19.636085141(2)\\
    $\dot{\nu}$ ($\times 10^{-10}\,\mathrm{Hz}\cdot\mathrm{s}^{-1}$) & $-2.521868(3)$ \\
    $\ddot{\nu}$ ($\times10^{-21}\,\mathrm{Hz}\cdot\mathrm{s}^{-2}$) & $4.6(1)$\\
    Epoch (MJD)& 60041.21699 \\
    Valid Range (MJD) & 59901--60318 \\
    Ephemeris & JPL-DE430 \\
    $\Delta\nu$ ($\times10^{-7}\,\mathrm{Hz}$) & $-1.042_{-0.074}^{+0.076}$ \\
    $\Delta\dot{\nu}$ ($\times10^{-15}\,\mathrm{Hz}\cdot\mathrm{s}^{-1}$) & $-7.4_{-6.1}^{+6.2}$\\
    $t_g$ (MJD) & $60132.158_{-4.633}^{+5.224}$ \\
    $|\Delta\nu/\nu|$ ($\times10^{-9}$) & $5.306_{-0.037}^{+0.038}$\\
    RMS residual (\SI{}{\micro\second}) & 829.8  \\
\hline
\end{tabular}
\caption{
    The inferred spin and glitch parameters. Errors refer to the 68\% confidence interval of the posterior distributions. }
\label{tab:spinparameters}
\end{table}

\begin{figure*}[ht]
    \centering
    \includegraphics[width=0.98\textwidth]{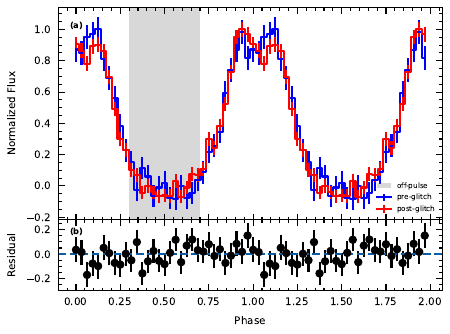}
    \caption{
    Cumulative pulse profiles before and after the glitch. \textbf{(a)} The pre-glitch pulse profile is represented by the blue line, while the red line indicates the post-glitch pulse profile. The grey area highlights the off-pulse range chosen as the background for normalization, within which each pulse profile has been background subtracted and normalized by dividing by the peak counts. \textbf{(b)} The residuals between the pre- and post-glitch pulse profiles. The error bars show 1 $\sigma$ uncertainties.}\label{fig:profile}
\end{figure*}

\subsection{Flux}
\label{sec:flux}

Looking for burst events or flux variations associated with the anti-glitch is an important clue for determining its, possibly internal, nature. To this end, we have examined the pulsed flux, burst activities, and variations in the pulse profile properties of~\target{}. 

Our analysis of pulsed flux relies on the NICER observations in the energy band 0.5-12\,keV. The background, identified as the un-pulsed phase, is selected for each pulse profile and is shown as the gray area in Figure~\ref{fig:profile}. The pulsed fluxes are normalized by the total number of Focal Plane Modules (FPMs) active during each NICER observation. These normalized results are displayed in the bottom panel of Figure~\ref{fig:phase}. The red squares in the bottom panel represent the averaged pulsed flux over time. We find no significant variation in the pulsed flux observed during the epoch of the anti-glitch event. 
On the other hand, we report the cumulative pulse profiles observed before and after the glitch in Figure~\ref{fig:profile}. Again, we find no significant changes in the pulse profiles before and after the anti-glitch.

We also employ the burst searching techniques described in \cite{cai2021search} for NICER light curves across several time scales, \SI{1}{\micro\second}, \SI{0.1}{\second}, \SI{1}{\second}, and \SI{10}{\second}. The energy band focused on is 2--8\,keV to limit the effects of possible particle background in the high energy band and the visible-light effect in the low energy band. This method is used to scrutinize NICER observations to identify X-ray bursts associated with the glitch event. We employ a time window to examine each time bin of the light curve, aiming to identify outliers exceeding background fluctuations with $3\,\sigma$ significance. The background fluctuations are based on Poisson statistics, determined by averaging the flux level in the intervals before and after the intervals of interest. There is no significant burst event discovered in the filtered NICER light curve. We also search the gamma-ray burst catalogue in the General Coordinates Network Circulars\footnote{https://gcn.nasa.gov/circulars}. No gamma-ray bursts were associated with \target{} during the period of glitch activity. The findings suggest that the anti-glitch we discovered in \target{} exhibits a radiatively quiet nature.

To summarize, we classify this event as an asymptotic anti-glitch, characterized by a negative post-glitch permanent frequency increment (see Figure~\ref{fig:modelz}). The glitch is radiatively quiet, exhibiting no significant variations in the shape of the pulse profile or pulsed flux, and no burst activities were observed in association with the glitch.

\section{discussion}\label{sec:discussion}


Glitches typically exhibit fractional amplitudes of $\Delta\nu/\nu\gtrsim10^{-9}$, coinciding with a decrease in $\dot\nu$. On the other hand, RPPs occasionally undergo `micro-glitches' with smaller amplitudes ($\Delta\nu/\nu\lesssim10^{-10}$, \citealt{chukwude2010observations,zhou2022pulsar}), manifesting as either positive or negative shifts in $\dot\nu$, albeit often evading detection \citep{2024Vahdat}. This could contribute to the observed rarity of anti-glitches in RPP.

Seven magnetars have been found exhibiting an anti-glitch or a spin-up glitch followed by a rapid spin-down resulting in a net spin-down effect: 1E 2259+586 \citep{younes2020radiatively, archibald2013anti, iccdem2012rxte}, 4U~0142+61 \citep{archibald2017swift, gavriil20112006}, 1E~1841-045 \citep{dib201416}, SGR~J1935+2154 \citep{ge2024spin, younes2023magnetar}, PSR~J1846-0258 \citep{livingstone2010timing, csacsmaz2014glitch}, SGR~J1900+14 \citep{woods1999variable}, and PSR~J1119-6127 \citep{dai2018peculiar, archibald2017swift}. 
Due to the sparse observations around the glitch epoch, it is crucial to determine whether the glitch observed in \target{} represents a true anti-glitch, or if the initial spin-up process was missed, leading the pulsar to over-recover to a net spin-down state, as in Figure~\ref{fig:modelz}. Thus, we extend our analysis further by applying a limiting case of the phenomenological model used by \citet{montoli2020bayesian} to our source, aiming to constrain the timing solution. 

\subsection{Glitch or anti-glitch?}
\label{sec:spinup}

To assess whether the glitch of \target{} is a regular spin-up glitch or an anti-glitch, we adopt a phenomenological model for the residuals based on a reduction of the one used for Vela's 2016 glitch \citep{montoli2020bayesian}, see Appendix~\ref{sec:app}. 
For $\delta t>0$ the phase residual model can be written as (cf. equation \ref{4parvel})
\begin{eqnarray}
\label{phasemodel}
\Delta\Phi(t)  =   \Delta\nu_{\infty}\, \delta t 
+ \frac{\Delta\nu_0 - \Delta\nu_{\infty}}{\lambda}\left( 1-e^{-\delta t \, \lambda} \right)
\end{eqnarray}
where $\Delta\nu_{\infty}$ is the asymptotic frequency offset and $\Delta\nu_0$ is the immediate post-glitch frequency jump. The values of $\Delta\nu_{\infty}$ and $\Delta\nu_0$ define different cases. For example, for spin-up glitches we can have that:
\begin{enumerate}
    \item $0<\Delta\nu_0<\Delta\nu_{\infty}$: A regular spin-up glitch with a fast initial jump and a following delayed spin-up, similar to what observed in the Crab~\citep{shaw+2018,ge2020discovery}.
    \item $0<\Delta\nu_{\infty}<\Delta\nu_0$: An overshooting glitch \citep{ashton2019rotational,pizzochero2020}. If the fast initial transient is not resolved, this resembles a regular spin-up glitch with a certain amount of recovery.
\end{enumerate}
The anti-glitch case is analogous, just the sign of the amplitudes is reversed, see Figure~\ref{fig:modelz}. Therefore, we now explore the parameter space for $\Delta\nu_{\infty}$ and $\Delta\nu_{0}$ in the phenomenological model, allowing them to be negative, to aptly describe the anti-glitch behavior. 
To this end, a similar MCMC approach is adopted, with the priors for each parameter detailed in Appendix~\ref{app4par}.
We find that $\Delta\nu_{0}$ -- the post-glitch frequency jump extrapolated at $t=t_g$ -- is broadly consistent with zero, $\Delta\nu_{0}=4.2_{-92.6}^{+115.4} \times10^{-9}\,\mathrm{Hz}$, and it is accompanied by a recovery timescale of $5.7_{-1.9}^{+5.0}\,\mathrm{days}$. Regarding the asymptotic frequency offset, the analysis points at an anti-glitch with $\Delta\nu_{\infty}=-1.08_{-0.09}^{+0.07}\times10^{-7}\,\mathrm{Hz}$. 
Given the large uncertainty on $\Delta \nu_0$, the Bayesian analysis suggests that this event is an \emph{asymptotic} anti-glitch, which could be preceded by a quickly recovered spin-up glitch or even by an undershoot, as shown in Figure~\ref{fig:modelz}. 
Unfortunately, continuous observation around the glitch epoch $t_g$ was unavailable, so we can not distinguish between these possibilities.  
In any case, it is worth noting that the recovery timescale for this event is shorter than that of the majority of other glitches observed in RPPs, at least those characterised by a single exponential recovery timescale~\citep{yu2013detection}.
There are only a few cases in the Crab Pulsar where the recovery timescales are as short as a few days \citep{wong2001observations}, and the post-glitch frequency does not fully recover to the pre-glitch solution. However, in our case, the small fractional change of $\Delta\dot{\nu}/\dot{\nu} \sim 2.7\times 10^{-5}$ is not sufficient to resolve an over-recovery scenario for the spin-up glitch in such a short recovery timescale.
In conclusion, the likelihood that the glitch event observed in \target{} represents a spin-up glitch appears to be marginal.


\subsection{Anti-glitch model}
\label{sec:model}

Our timing analysis of \target{} reveals a radiatively quiet anti-glitch in an RPP. 
To our knowledge, this is the first study to report such an occurrence.

There are several proposed models for the anti-glitches in magnetars, but are not suitable for RPPs. 
For example, \cite{garcia2015simple} and \cite{mastrano2015interpreting} interpreted anti-glitches as a result of the decay of its internal toroidal magnetic field component, which transforms a stable prolate configuration into an unstable one. Unlike magnetars, RPPs do not possess sufficiently strong magnetic fields to form a prolate configuration; instead, the oblate configuration of RPPs \citep{baym1971neutron} tends towards a more spherical configuration. Thus, when a glitch occurs in RPPs, the moment of inertia decreases, leading to a spin-up glitch~\citep{ruderman1969,baym+1969}. 

\cite{tong2014anti} posited that the particle wind contributes to the observed flux enhancement in \oneE{} and exerts a net spin-down torque on the NS. Alternatively, \cite{huang2014anti} speculate that anti-glitches may result from collisions of a pulsar with a small body, where gravitational potential energy is released either in a short hard X-ray burst or in a soft X-ray afterglow. However, our results on \target{} do not indicate any significant radiative change that can be associated with the anti-glitch event.

Recently, \cite{yim2024continuous} established a toy model based on ejecta from the magnetosphere that can lead to glitch or anti-glitch events. 
According to this model, some mass  $M_0$ can be expelled from the magnetic pole but may get trapped as it travels outside through the magnetosphere. 
According to their analysis (refer to Figures 3 and 4 therein), the ejecta results in an anti-glitch depending on two conditions:

the magnetic inclination angle from which ejecta are expelled must be large, and the ejecta must be expelled to a significant distance, at least $10 R_0$, where $R_0$ is the NS radius. 
For an NS like \target{}, the co-rotation radius $R_{\mathrm{co}}$ is 
\begin{equation}
\frac{R_{\mathrm{co}}}{R_0}\approx 170\left[\frac{M}{1.4\,M_{\odot}}\right]^{\frac{1}{3}} \!
\left[\frac{\nu}{1\,\mathrm{Hz}}\right]^{-\frac{2}{3}} \!
\left[\frac{R_0}{10\,\mathrm{km}}\right]^{-1}
\! \! \!\approx 23,
\end{equation}
which is at the edge of the anti-glitch condition. This elucidates both the possibility and the rarity of anti-glitch events, indicating that, for such an event to occur, the ejecta must be expelled to a distance nearing the co-rotation radius. However, this toy model posits that the motion of the trapped ejecta induces free precession, leading to modulations in the pulse profile, polarization, and timing. 
Since we have not observed variations in the X-ray band, additional observational evidence is required to establish this scenario as a dominant mechanism underlying the anti-glitch phenomenon.

Regarding the possible internal origin, \cite{kantor2014anti} developed an anti-glitch model that extends the standard scenario of pulsar glitches. Glitches in RPPs are interpreted as a momentum transfer between the superfluid and the normal components in the inner crust or outer core, depending on where the quantized vortices can pin~\citep{Antonelli2022}. 
\citet{kantor2014anti} extended this framework by considering that the superfluid fraction also depends on the superfluid current (it increases with decreasing velocity between the normal and superfluid components). This results in a mass redistribution between the superfluid and normal part of an NS when the velocity lag between the two components changes after the quantized vortices unpin.
Following the notation of \citet{kantor2014anti}, the total angular momentum conservation reads
\begin{equation}
\label{eq:angularmomentumconservation}
    I_{\mathrm{c0}}\Omega_{\mathrm{c0}} + I_{\mathrm{s0}}\Omega_{\mathrm{s0}} = I_{\mathrm{c1}}\Omega_{c1} + I_{\mathrm{s1}}\Omega_{\mathrm{s1}} \, ,
\end{equation}
where $I$ denotes the moment of inertia, $\Omega$ represents the angular velocity, and the subscripts `c' and `s' refer to the normal (i.e., everything that is non-superfluid) and superfluid components, respectively, with `0' indicating pre-glitch and `1' post-glitch parameters. Taking into account the changes in the moments of inertia, $I_{\mathrm{c}}'$ and $I_{\mathrm{s}}'$ due to mass transfusion between the components ($I_{\mathrm{c}}'+I_{\mathrm{s}}'=0$), the change in angular velocity of the normal component after the glitch is 
\begin{equation}
\label{eq:kantorglitch}
    \delta\Omega_{c}=
    -\frac{I_{s0}- I'_{c}\Delta\Omega_{0}}{I_{c0}+I'_{c}\Delta\Omega_{0}}\delta\Omega_{s} \, ,
\end{equation}
where $\Delta\Omega_{0}=\Omega_{\mathrm{s0}}-\Omega_{\mathrm{c0}}>0 $ is the angular velocity lag before the unpinning and $I_{s,c}'=dI_{s,c}/d\Delta\Omega$ is the rate of change of the partial moment of inertia with the lag $\Delta\Omega=\Omega_s-\Omega_c$.
This gives an anti-glitch ($\delta\Omega_{c}<0$) when $|I'_{s}|\Delta\Omega_{0} > I_{s0}$, a condition that can be met depending on two physical parameters: the stellar temperature and the rotational lag $\Delta\Omega$. The higher the internal temperature ($T\gtrsim 10^7\,\mathrm{K}$) 
and the lag before vortex unpinning ($\Delta \Omega\gtrsim 1\,$rad/s), the more likely is to meet the anti-glitch condition (see Figures 2 and 3 in \citealt{kantor2014anti}). 
\target{} is arguably one of the youngest pulsars and is a possible candidate to satisfy the anti-glitch criterion. First, the internal temperature in young pulsars like \target{} is expected to be relatively high, $T\gtrsim 10^8$K. Moreover, the increase in the spin-down rate in 2011 \citep{marshall2015discovery, ge2019brightening} might have boosted the lag $\Delta\Omega$ in \target{}. 
It is possible to provide an upper limit to the increase of $\Delta\Omega$ since the last reported glitch, around MJD 52927 \citep{ferdman2015long}.
The long-term spin-down rate changed from $-1.87\times10^{-10}\,\mathrm{Hz/s}$ to $-2.52\times10^{-10}\,\mathrm{Hz/s}$ after a spin-down rate change at MJD 55866 \citep{ge2019brightening}. 
Therefore, the maximum variation of the lag achieved during this time interval is about $\delta \Omega \approx 0.88\,$rad/s, assuming no undetected glitches have occurred and the superfluid's angular velocity was conserved between the last reported glitch and our anti-glitch event (i.e., vortices were always pinned). 

    This value represents an upper limit to how much $\Delta\Omega$ could have changed from MJD 52927 to $t_g\approx$ MJD~60134. Its order of magnitude is consistent with the typical values of $\Delta\Omega_0\sim 2$\,rad/s needed to trigger an anti-glitch for a star whose internal temperature is $T\sim 10^8$\,K. This seems to indicate that triggering an anti-glitch is indeed possible, considering that it is unlikely that the previous glitch at MJD 52927 would have emptied the entire superfluid's momentum reservoir. 

This anti-glitch model is likely better suited for magnetars, as it requires a strong toroidal magnetic field to pin vortices in the outer core \citep{alpar2017}: vortex unpinning in the outer core would result in an anti-glitch because the critical temperature for superfluid neutrons is significantly lower there than in the inner crust, allowing for a more efficient matter transfusion from the two components.

    Thus, even though \target{} is not classified as a magnetar, its internal toroidal component would have to be much stronger than its poloidal field, but this is not necessarily a problem. The Bayesian analysis of Vela's 2016 glitch also provides a first observational clue for pinning in the outer core due to a strong magnetic field~\citep{montoli2020bayesian}, a scenario often invoked in pulsar glitch models~\citep[e.g.,][]{gerci2014ApJ,sourie2020}.
Furthermore, magnetohydrodynamic simulations reveal that it is possible to develop stable internal toroidal magnetic fields that are relatively high with respect to the poloidal component \citep{ciolfi2013mnras,ciolfi2014AN, lander2014contrasting}. Therefore, given the current observational results, we cannot rule out the anti-glitch scenario of \citep{kantor2014anti} for~\target{}.

Finally, we note that \citet{kantor2016} also proposed a scenario for the evolution
of rapidly rotating NSs in low-mass X-ray binaries. In such objects, a resonant interaction of normal r-modes with superfluid inertial modes produces new stable regions in the $T$-$\nu$ plane (again, $T$ is the internal temperature). 
The spin evolution of these hot and rapidly rotating NSs moves around the boundary between stable and unstable regions, periodically crossing it.
The periodic excitation of r-modes implies the possibility of anti-glitches in these objects, whose frequency jumps unfold over a time scale that varies, depending on the system, from hours to months.
Although this model can be applied to spinning down pulsars, \target{} is located in a different region of the $T$-$\nu$ plane, where r-modes are expected to be stable. However, future observations of anti-glitches in \target{} may reveal the extent of their periodicity. In fact, while for glitches (or anti-glitches) in the standard vortex scenario, it is natural to expect no periodicity (vortex unpinning is a multi-threshold phenomenon, see the discussion in \citealt{Antonelli2022}), the r-mode scenario relies on a single and global instability threshold, resulting in a periodic behaviour.

In conclusion, our Bayesian analysis points to an asymptotic anti-glitch event in an RPP, marking the first occurrence of such a phenomenon observed beyond magnetars and accreting pulsars. 
The radiatively quiet nature associated with this glitch event suggests that internal processes within the neutron star are responsible for this anti-glitch. To date, the possible physical mechanisms that can lead to anti-glitches are based on magnetar properties, as they were the only isolated neutron stars that exhibited anti-glitches. We speculate that the anti-glitch of \target{} can be explained through the transfer of mass from the superfluid to the normal component, as suggested by \citep{kantor2014anti} under the condition that the internal toroidal field of the pulsar is significantly higher than its dipole fields, a possibility that may also be met in the Vela pulsar \citep{gerci2014ApJ,montoli2020bayesian}. 
Further observations on timing and multi-wavelength monitoring, together with a more systematic analysis of proposed models for anti-glitches, are needed to understand these phenomena in RPPs.

\begin{acknowledgements}
The authors would like to express our sincere gratitude to the
referee for their insightful comments. This work is supported by the National Key R\&D Program of China (2021YFA0718500) from the Minister of Science and Technology of China (MOST). The authors thank the support from the National Natural Science Foundation of China (Grants 12373051, 12333007) and the International Partnership Program of the Chinese Academy of Sciences (Grant No. 113111KYSB20190020). This work is supported by National Key R\&D Program of China (2023YFE0117200) and National Natural Science Foundation
of China (Grant 12373041). We acknowledge the use of data from the \emph{Neutron star Interior Composition Explorer} (NICER) mission, a project led by the NASA/Goddard Space Flight Center. This research also uses data obtained via the High Energy Astrophysics Science Archive Research Center Online Service, provided by the NASA/Goddard Space Flight Center. The authors acknowledge NASA’s Astrophysics Data System (ADS) Bibliographic Services and the arXiv repository. 
We are grateful for the insightful discussions with Dr.~Abdujappar Rusul, Dr.~Ce Cai, and Prof.~Lin~Lin. 
\end{acknowledgements}



\software{
    astropy \citep{robitaille2013astropy},  
    TEMPO2 \citep{hobbs2006tempo2}, 
    Stingray \citep{huppenkothen2019stingray},
    TAT-pulsar \citep{tuo2022orbit}.  
    }



\newpage

\appendix
\section{Phenomenological model for the glitch/anti-glitch residuals}
\label{sec:app}

For completeness, we discuss the parameter space of the phase residue model in equation \eqref{phasemodel}, and show that it can be recovered as a limit of the one used for the fit of Vela's 2016 glitch~\citep{ashton2019rotational,montoli2020bayesian}. 

\subsection{Five-parameter model for overshooting glitches and anti-glitches}

\begin{figure*}[ht]
    \centering
    \includegraphics[width=0.99\textwidth]{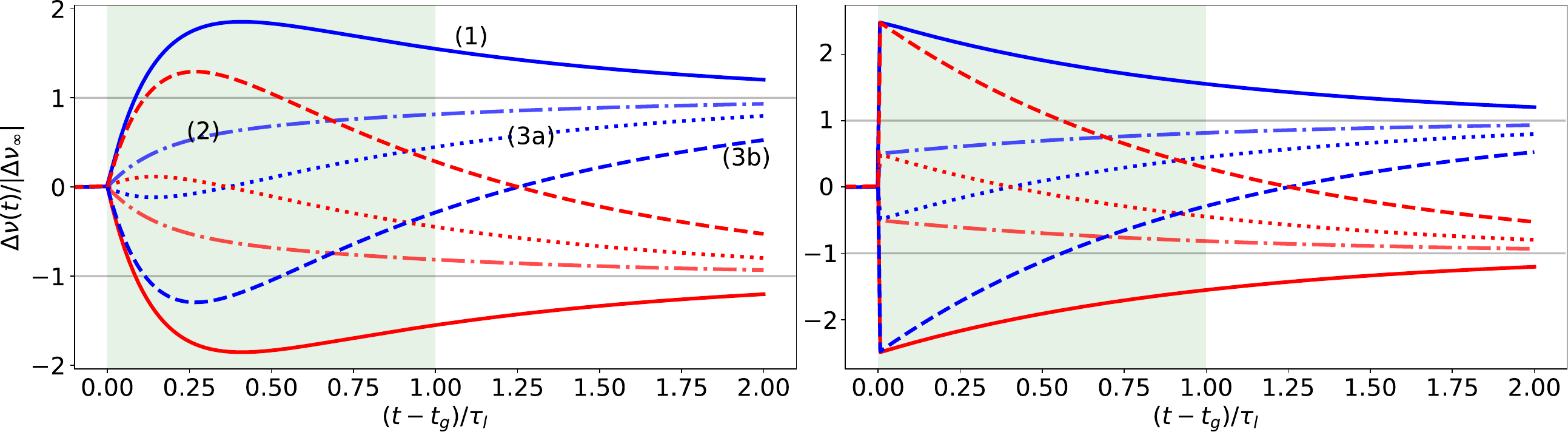}
    \caption{
    Sketch of the possible realizations of model \eqref{eq:dnup} (left panel, $\tau_s=\tau_l/7$) and model \eqref{4parvel} (right panel, $\tau_s=0$). The blue curves represent events that are `asymptotic spin-up' events ($\Delta\nu_\infty>0$), and the red ones are `asymptotic spin-down' events ($\Delta\nu_\infty<0$). The solid curves refer to case 1: $\omega=2.5$ (overshooting glitch/undershooting anti-glitch). The dash-dotted curves refer to case 2: $\omega=0.5$ (two-timescale spin-up/spin-down). Case 3 is represented by the dotted (3a, $\omega=-0.5$) and dashed (3b, $\omega=-2.5$) curves. The left panel contains the same curves in the limit $\tau_s=1/\lambda_+=0$.
    }\label{fig:modelz}
\end{figure*}

Assume that $\nu_{pre}(t)$ and $\nu_{post}(t)$ describe the pulsar evolution before and after the glitch/anti-glitch epoch, that is conveniently split at $t_g=0$.
The rotation frequency $\nu(t)$ and its residue $\Delta\nu(t)$ for any time $t$ are 
\begin{equation}
 \nu(t) = \theta(t) \, \nu_{post}(t) +  \theta(-t) \, \nu_{pre}(t) = \nu_{pre}(t) + \Delta \nu(t) 
  \qquad \qquad
 \Delta \nu(t) = \theta(t) \left( \nu_{post}(t) - \nu_{pre}(t) \right) \, ,
\end{equation}
where $\theta(t)$ is the Heaviside step function. Now, we have to choose a suitable phenomenological form for $\Delta \nu(t)$. 
The one used for the Bayesian fit of Vela's 2016 glitch is~\citep[cf. equation 4 of][]{montoli2020bayesian}
\begin{equation}
\begin{split}
& \Delta \nu(t) 
\, = \, 
\theta(t) \, 
\Delta\nu_\infty \left[ 1- \omega \, e^{-t \lambda_+} - (1-\omega)\,  e^{-t \lambda_-}  \right]
\\
& \Delta \Phi(t) = \theta(t) \int_0^t \Delta \nu(t') dt' \, 
        =\theta(t) \, \Delta\nu_\infty 
        \left[ t + 
        \dfrac{\omega }{\lambda_+}  \left(e^{-t \lambda_+}  -1\right) +
        \dfrac{1-\omega }{\lambda_-} \left(e^{-t \lambda_-}  -1\right) 
        \right]  
\end{split}
\label{eq:dnup}
\end{equation}
where $\omega$ is a dimensionless parameter that sets the overall behaviour of the glitch 
(see Figure~\ref{fig:modelz}), $0<\lambda_-<\lambda_+$ are the inverse of two recovery timescales (the short one given by $\tau_s=1/\lambda_+$ and a longer one, $\tau_l=1/\lambda_-$) and  $\Delta \nu_\infty$ is the asymptotic glitch/anti-glitch amplitude. We have `asymptotic spin-up' events for $\Delta\nu_\infty>0$, and `asymptotic spin-down' events for $\Delta\nu_\infty<0$.
The model in equation \eqref{eq:dnup} allows fitting situations where the normal component overshoots\footnote{
    \rlap{\parbox{\dimexpr\textwidth-20pt\relax}{
    Given a glitch triggered at $t=0$, assume that there a is a time $t_{\rm max}>0$ such that $\Delta \dot{\nu}(t_{\rm max})=0$. A spin-up glitch `overshoots' if $\Delta {\nu}(t_{\rm max})$ is a maximum, $\Delta \ddot{\nu}(t_{\rm max})$<0. Conversely, an anti-glitch `undershoots' if $\Delta {\nu}(t_{\rm max})$ is a minimum, $\Delta \ddot{\nu}(t_{\rm max})$<0.
    }}},
which is realised when~\citep[cf. equation 5 of][]{montoli2020bayesian}
\begin{equation}
        \frac{t_{\rm max}}{  \tau_l - \tau_s } \, = \,   
\log \left( \frac{\tau_l}{\tau_s} \right) 
+
\log \left( \frac{ \omega}{\omega-1} \right)
\, >0 \, .
\label{eq:tmax}
\end{equation}
The time $t_{\rm max}$ is a real positive number only when $\omega <0$ or $\omega > 1$. 
All the possible qualitative behaviours of \eqref{eq:dnup} are sketched in Figure~\ref{fig:modelz}:
\begin{enumerate}
    \item $\omega>1$: Overshooting glitch ($\Delta\nu_\infty>0$) or undershooting anti-glitch ($\Delta\nu_\infty<0$).
    \item $0<\omega<1$: Two-timescale spin-up/spin-down, depending on the sign of  $\Delta\nu_\infty$. The case $\Delta\nu_\infty>0$ resembles the `delayed spin-up' behaviour observed during the spin-up phase of Crab's 2017 glitch~\citep{shaw+2018}.
    \item $\omega<0$: In this case, the glitch/anti-glitch distinction depends on whether one looks at the short-timescale or the long-timescale behaviour. 
    Since this is the most exotic possibility, in Figure~\ref{fig:modelz} we provide two examples for this case (3a, $-1<\omega<0$ and 3b, $\omega<-1$), that are not qualitatively different. 
\end{enumerate}
When \eqref{eq:dnup} is used in a fit procedure, we have to consider that the glitch/anti-glitch epoch $t_g$ is an extra parameter that should be inferred together with the other parameters: $\omega \in \mathbb{R}$, $t_g \in \mathbb{R}$, $\Delta\nu_\infty \in \mathbb{R}$, $0 < \lambda_- < \lambda_+$ (or, equivalently, $ 0<\tau_s<\tau_l$), for a total of 5 real parameters to be extracted from the timing data.

\subsection{Unresolved short timescale: four-parameter model}
\label{app2}

Assume that the timescale $\tau_s$ can not be resolved with the available timing data. 
The limit $\tau_s\ll\tau_l$ gives:
\begin{equation}
\label{4parvel}
\Delta \nu(t) 
\, = \, 
\theta(t) \, 
\Delta\nu_\infty \left[ 1 - (1-\omega)\,  e^{-t \lambda_-}  \right] 
        \qquad \quad
\Delta \Phi(t) = \theta(t) \Delta\nu_\infty 
        \left[ t +
        \dfrac{1-\omega }{\lambda_-} \left(e^{-t \lambda_-}  -1\right) 
        \right]
\end{equation}
Hence, we have an instantaneous jump of amplitude $\Delta\nu_0=\omega \Delta\nu_\infty $ at $t=t_g$, that then relaxes to $\Delta\nu_\infty$ for $t \gg \tau_l$.
The model \eqref{4parvel} is formally equivalent to the one of \citet{baym+1969}, see \cite{Antonelli2022} for a revised discussion of the original model. This can be seen by introducing the `healing parameter' $Q$,
\begin{equation}
\label{4parvelllll}
\Delta \nu(t) 
\, = \, 
\theta(t) \, 
 \Delta\nu_0\left[  1- Q \left(1-  e^{-t \lambda_-} \right) \right] \qquad \quad Q=(\Delta\nu_0-\Delta\nu_\infty)/\Delta\nu_0 = (\omega-1)/\omega \, .
\end{equation}
Typically, $0<Q<1$ is observed \citep{Crawford2003ApJ}: $Q=0$ is a step-like glitch with no detectable relaxation, and $Q=1$ is a glitch with complete relaxation. Since the condition $0<Q<1$ is equivalent to $\omega>1$, the simplest model to fit glitches formally corresponds to the $\tau_s=0$ limit of an agnostic model for overshooting glitches. Moreover, as long as \eqref{4parvelllll} is regarded as an agnostic model, we do not necessarily have to stick to the parameter space of typical spin-up glitches, and we can assume that $Q\in \mathbb{R}$. 
Despite its simplicity, the model \eqref{4parvel} is already quite rich as it can represent all the possibilities sketched in Figure~\ref{fig:modelz}. Hence, it provides a reasonably simple prescription to fit situations where one is unsure about the event's glitch/anti-glitch nature, as in our case.  

\subsection{Agnostic model for the Bayesian fit }
\label{app4par}

The Bayesian fit is performed by considering all the timing data available before and after the (unknown) glitch epoch $t_g$. For the frequency residues model, we adopt the form \eqref{4parvel} but we also extend it to account for the possibility of a permanent change of the frequency derivative:  
\begin{equation}
    \Delta\nu(t) =\theta(\delta t) \left[ 
  \Delta \nu  +  \Delta\dot{\nu} \, \delta t  + 
    \Delta\nu_{d} \, e^{-\delta t/\tau}    
    \right] \, ,
\end{equation}
cf. with equation \eqref{4parvelllll} by using 
$\Delta\nu=\Delta\nu_{\infty}$ (the `permanent' frequency jump is just the `asymptotic' frequency jump), $\Delta\nu_d=\Delta\nu_0-\Delta\nu_\infty$, $Q=\Delta\nu_d/\Delta\nu_0$, $\tau=1/\lambda_-$. Therefore, the full model that we have to fit is $\nu(t)=\nu_{pre}(t)+\Delta\nu(t)$. Since we have to fit the ToA, in practice we fit the corresponding phase model, namely $\Phi(t) = \Phi_{pre}(t)+\Delta\Phi(t)$, where $\Phi_{pre}(t)$ is exactly given in \eqref{eq:phi} and
\begin{equation}
\label{final}
    \Delta\Phi(t) = \theta(\delta t) \left[  
    \Delta\nu \, \delta t + \Delta \nu_d \, \tau \left( 
    1 -e^{-\delta t/\tau}
    \right) +\frac{\Delta\dot{\nu}}{2}\delta t^2
    \right]\, ,
\end{equation}
for a total of 8 real parameters (i.e., $\nu$, $\dot\nu$, $\ddot\nu$, $t_g$,  $\Delta\nu$, $\Delta\dot{\nu}$, $\Delta\nu_d$, $\tau$) to be extracted from the timing data.
For these parameters, we take flat prior distributions within reasonably large intervals: 
\begin{eqnarray}
   \label{priors}
& \nu \in [19.63,19.64]\, \text{Hz} \qquad 
 \dot\nu \in [-2.6,-2.5]\times10^{-10}\, \text{Hz/s} \qquad
 \ddot{\nu} \in [4.3\times 10^{-24},4.3\times 10^{-19}]\, \text{Hz/s}^2  \qquad
 \nonumber  \\
& \Delta{\nu} \in [-1,1]\times10^{-4}\, \text{Hz} \qquad
 \Delta{\dot{\nu}} \in [-1,1]\times10^{-12}\, \text{Hz/s}^2 \qquad
 \Delta{\nu}_d \in [-1,1]\times10^{-4}\, \text{Hz} \qquad
\\
& \tau \in [0,200]\, \text{day} \qquad
 t_g \in [60120, 60150]\, \text{MJD}.  \nonumber       
\end{eqnarray}
\bibliography{bibliography}{}
\bibliographystyle{aasjournal}
%
\end{document}